\documentstyle[11pt,aaspp]{article}
\tighten
\journalid{}{}
\articleid{}{}
\slugcomment{}
\begin{document}

\title{DIFFUSE IONIZED GAS IN SPIRAL GALAXIES:
PROBING LYMAN CONTINUUM PHOTON LEAKAGE FROM HII REGIONS?}

\author{Annette M. N. Ferguson\altaffilmark{1} and Rosemary F. G. Wyse\altaffilmark{1,2}}
\affil{Department of Physics and Astronomy, The Johns Hopkins University,
 Baltimore, MD 21218}
\author{J. S. Gallagher III}
\affil{Department of Astronomy, University of Wisconsin, 
Madison, WI 53706}
\author{Deidre A. Hunter}
\affil{Lowell Observatory, 1400 W. Mars Hill Rd., Flagstaff, AZ 86001}
\altaffiltext{1}{Visiting Astronomer, Cerro Tololo Inter-American Observatory. 
CTIO is operated by AURA, Inc.\ under contract to the National Science
Foundation.} 
\altaffiltext{2}{Center for Particle Astrophysics, University of California,
Berkeley, CA 94720, USA}
\begin{abstract}
As part of a large study to map the distribution of star formation
across galactic disks, we have obtained deep H$\alpha$ images of the
nearby Sculptor Group spirals NGC~247 and NGC~7793.  These images are
of sufficiently high quality that they allow identification and
analysis of diffuse H$\alpha$ emission at surface brightness levels
ranging from those of extremely low density HII regions to those of the
local Galactic disk diffuse emission. This paper presents a study of
the large scale distribution and global energetics of diffuse ionized
gas (DIG) in these galaxies and investigates the association between
DIG and discrete HII regions.   Our results support the hypothesis that
the DIG is photoionized by Lyc photons which leak out of traditional
HII regions, and suggest that the local HI column density plays a role
in regulating the amount of leakage which can occur.   This
interpretation has profound implications for the derivation of
star-formation rates based on H$\alpha$ emission--line fluxes since HII
region counts alone will lead to significant underestimates of the true
rate.  The contribution of the diffuse H$\alpha$ component to the total
H$\alpha$ emission, ie. the diffuse fraction, in these galaxies is
found to be similar to values found in other disk galaxies with
differing Hubble types and star formation rates.  The constancy of the
diffuse fraction is rather unexpected and implies that the overall
fraction of photons which can leak out of HII regions and ionize the ISM
over large scales is relatively invariant from one galaxy to another.
\end{abstract}

\section{Introduction}

Understanding the relationship between the star formation process and
the interstellar medium is a key step towards unravelling the complex
history of galaxy evolution.  A longstanding problem has been the
nature and importance of the feedback process by which massive stars
deposit energy into the interstellar medium via photoionization,
stellar winds and supernovae.  This feedback mechanism affects the
physical and dynamical state of the interstellar medium and hence
influences the rate and distribution of subsequent star formation.   OB
stars clearly provide the largest source of Lyman continuum (Lyc)
photons in typical spiral galaxies (Abbott 1982); a major issue
concerns whether the bulk of these photons are deposited over localized
regions, such as the Str\"omgren spheres which define traditional  HII
regions,  or whether a significant fraction can escape from the regions
of recent star formation where they were created and can ionize the
interstellar medium over much larger scales.    If large--scale
ionization due to Lyc photon leakage from HII regions were a common
feature of spiral galaxies, then the resulting diffuse ionized gas
(DIG) component would provide an important tracer of the feedback
process due to star formation, as well the structure of the
interstellar medium.  It would be of crucial importance to include this
component in calculations of radial and global star formation rates
derived on the basis of H$\alpha$ emission line fluxes.  Furthermore,
the existence of such a DIG component would have direct bearing on the
energy balance of the interstellar medium, regardless of the source of
photons that maintain it in an ionized state.
\par
 Indeed, the existence of widespread diffuse ionized gas lying outside
the boundaries of traditional HII regions has been known for over 20
years (Reynolds {\it et al.} 1971;  Monnet 1971), but
the source of this ionization remains the subject of much debate.
Pulsar dispersion measures and optical line emission have been used to
trace the Galactic DIG component (the `Reynolds Layer') and have
established it to be an extremely important component of the
interstellar medium, occupying more than 20\% of the interstellar
volume, contributing at least $\onethird$ of the total HI column at the solar
circle and constituting $\sim$~90\% of the ionized hydrogen mass in the
Galaxy (see review by Reynolds 1991 and references therein).  The
location of the Sun close to the plane of the disk of the Milky Way
means that the vertical structure of the ionized gas is more easily
studied than is its radial distribution; it has been established that
the Galactic DIG is distributed in a thick disk with a scale--height
locally of $\sim$~1~kpc.  Much of the DIG (sometimes referred to as the
Warm Ionized Medium; WIM) is observed to be in the form of discrete
structures, such as loops, filaments and shells (which are not
obviously associated with discrete HII regions) while the remainder is
in the form of an apparently unstructured, diffuse background (Reynolds
1993).    The power required to maintain this low-density ($n_e
\sim$~0.1~cm$^{-3}$), warm ($<$~10000~K) ionized component is large --
1.0 $\times$~10$^{-4}$~erg~s$^{-1}$~cm$^{-2}$ of disk  -- and can be
met easily only by ionizing photons from OB stars, although the kinetic
energy from supernovae could possibly suffice (Reynolds 1984; Kulkarni
and Heiles 1988).
\par
The identification of OB stars as the source of the ionization of the
Galactic DIG would require that Lyman continuum photons be able to
travel several hundred parsecs both perpendicular to the plane, to
account for the derived vertical scale--height of a kiloparsec, and
parallel to the plane, to account for the DIG seen at large radial
distances from OB stars (see Reynolds 1990).  These requirements are
even greater in irregular galaxies, where ionized gas has been observed
at distances of $\gtrsim$~1~kpc from bright OB associations (Hunter and
Gallager 1990; Hunter and Gallagher 1992).  Recent calculations suggest
that it is indeed possible for a large fraction of ionizing photons to
penetrate sufficiently large distances from their point of origin,
depending on cloud properties and distributions, as well as on their
radial gas distributions (Dove and Shull 1994; Miller and Cox
1993).  The observed optical emission line ratios of the Galactic DIG
can be explained by a rather dilute radiation field and hence
argue further for photoionization by distant OB stars (D\"omgorgen and
Mathis 1994).  Alternative sources for the ionizing radiation that
maintains the DIG which have been proposed include shocks, decaying
massive neutrinos (Sciama 1990), turbulent mixing
layers (Slavin, Shull and Begelman 1993) and Galactic microflares
(Raymond 1992).   In addition, significant contributions to the 
ionizing flux may come from evolved stellar objects such as planetary
nebula nuclei, hot white dwarfs and blue horizontal branch stars.

\par
The large scale radial distribution of the DIG across galactic disks
can provide stringent constraints on the source of its ionization.  For
example, if the amount of Lyc photons produced in star--forming regions
were the only factor responsible, then the DIG distribution and
intensity should be expected to correlate, over both small and large
scales,  with that of discrete HII regions.  Furthermore,  DIG
properties should also be expected to vary systematically with the
morphological type of the host galaxy, reflecting the underlying
variation in star formation and structure within the ISM (Kennicutt,
Edgar and Hodge 1989).  Study of such correlations requires
observations of external galaxies at moderate inclinations. To date,
few quantitative studies of the DIG in external galaxies have been
carried out, and the emphasis has largely been on edge--on systems
(eg.  Rand, Kulkarni and Hester 1990; Dettmar 1992; Veilleux {\it et
al.} 1995).  Walterbos and Braun (1994; hereafter WB94) studied the DIG
component in selected areas of the nearby spiral M31, but their small
field of view restricted them to local analysis and they could not
assess the overall distribution and energetics of the ionized gas they
detected.
\par
The first quantitative study of the large--scale distribution and
global energetics of diffuse ionized gas in moderately face--on
external disk galaxies is presented in this paper.  Our data consist of
large field of view, deep H$\alpha$ images of Sculptor Group late--type
spirals NGC~247 and NGC~7793 which were obtained as part of a large
study to map the distribution of star formation across galactic disks.
NGC~7793 has been previously noted for having a significant DIG
component (Monnet 1971) but this earlier work was limited to
photographic data, with surface brightness and spatial resolution
limits significantly lower than the CCD data presented here.  Basic
properties of these galaxies are listed in Table 1.   In the present
paper, we derive the radial and azimuthal distributions of the DIG and
investigate the association between DIG and discrete HII regions.  We
also study the global and radial variation of the diffuse fraction and
the Lyc power requirements, which allow tight
 constraints to be placed on the origin and nature of DIG component in
 spiral galaxies.  Our results support the hypothesis that the DIG is
photoionized by Lyc photons which leak out of discrete HII regions, and
suggest that the local HI column density plays a role in regulating the
amount of leakage which can occur.   Given this hypothesis, we discuss
the great importance of taking account of the DIG component when
deriving star formation rates based on H$\alpha$ emission line fluxes.

\section{Observations and Data Analysis} 

\setcounter{footnote}{0}

The observations were carried out at the Cerro Tololo Inter--American
Observatory (CTIO) during the nights of December 14 -- 18, 1993 and
September 28 -- October 2, 1994.  Images were obtained with the Tek
2048 $\times$ 2048 CCD on the 1.5m telescope, at f/7.5.  The resulting
pixel scale was 0.44$''$ and the field of view was 15$'$ on a side,
corresponding to physical sizes of approximately 6~pc and 13~kpc,
respectively, at the mean distance of our galaxies.  The H$\alpha$
filter was 68 {\AA} wide (FWHM) centered on 6554 {\AA} and encompassed
the [NII] lines.  A  broadband R filter was used for the continuum
observations.   Total exposure times ranged from 3600 -- 5700 secs in
H$\alpha$ and from 600 -- 840 secs in R.  The seeing was in the range
of 1.2   -- 2.5$''$.
\par
The large angular extent of the galaxies on the sky -- 25th magnitude
isophotes of 10.1$'$~$\times$~6.1$'$ for NGC~7793 and
5.4$'$~$\times$~19.9$'$ for NGC~247 (Carignan 1985) --resulted in
several fields being required to map each galaxy completely.  Using the
`shift and stare' technique, we imaged each galaxy in two sections
along the major axis.   Well--exposed, median--filtered twilight sky
frames were used to flat-field the images, and all frames were
corrected to an airmass of zero.  Images of each section were
registered and those taken through the same filter were combined using
an average sigma-clipping algorithm.   The sky to be subtracted was
determined in each of these combined H$\alpha$ and R--band frames as
the mean of the mean pixel value in a series of 100~ $\times$~100 pixel
boxes positioned well outside the Holmberg diameters.  The uncertainty
in the determination of the sky value due to large scale flat--fielding
errors was less than or equal to 1\% in each case.    Foreground stars
in the frames were used to determine the scaling factors between the
H$\alpha$ and R--band images, and the scaled continuum images were
subtracted from the H$\alpha$.
\par
Observations of standard stars from the list of Stone and Baldwin
(1983) were used to calibrate the H$\alpha$ images. Instrumental
magnitudes of several stars were measured on the H$\alpha$
continuum--unsubtracted images.  Standard magnitudes were derived from
Stone and Baldwin (1983) by integrating the interpolated flux at
H$\alpha$ across the filter band--pass, assuming a top--hat
transmission curve.  Comparison of the two magnitudes provided the
conversion between observed counts and absolute flux.  There are no
previously measured HII regions with which to compare our
measurements.  Our neglect of factors such as the true shape of the
filter transmission curve and the variation of stellar flux across the
band--pass is likely to lead to uncertainties in the calibration factor
of $\pm$~20\%.  The average sensitivity of the H$\alpha$
continuum-subtracted images, taken to be 1$\sigma$ of the sky
background, is $4.3 \times 10^{-18}$~erg~s$^{-1}$~cm$^{-2}$~pix$^{-1}$
for the NGC~7793 image, and $6.4 \times
10^{-18}$~erg~s$^{-1}$~cm$^{-2}$~pix$^{-1}$ for the NGC~247 image.
These values correspond to emission measures{\footnote{Emission measure
is related to Rayleighs, the commonly used unit of surface brightness
in Galactic DIG studies, by EM (pc cm$^{-6}$) =
2.78$\times$I$_{H\alpha}$ (Rayleighs) for T$_{e}$=10$^4$ K.} (EM) per
pixel of $\sim 11$~pc~cm$^{-6}$ and $\sim 16$~pc~cm$^{-6}$
respectively, for an assumed electron temperature of 10$^4$~K .   The
scaled R--band images had sufficiently high  S/N that the noise in the
H$\alpha$ continuum--subtracted images was limited by the noise in the
original H$\alpha$ frame alone.
\par
The two separate H$\alpha$  continuum--subtracted images of each galaxy
were mosaiced together to give a field of 13.8$'$~$\times$~17.5$'$
centered on NGC~7793 and 10.6$'$~$\times$~23.8$'$ centered on NGC~247.
In the mosaicing process, care was taken to ensure that fluxes of
objects common to both frames were matched.  The sky levels across the
mosaiced images were measured in areas beyond the Holmberg diameters
and were 0~$\pm$~1.5~pc~cm$^{-6}$ in the NGC~7793 frame and  0~
$\pm$~3.2~pc~cm$^{-6}$ in the NGC~247 frame.  This combination of low
sky noise, small flat--fielding errors and a large number of pixels
allows us to reach very low H$\alpha$+[NII] surface brightnesses,
comparable to those observed locally in the Reynolds layer ($\sim
6$~pc~cm$^{-6}$; Reynolds 1984).  Thus,  a direct comparison of the
diffuse ionized ISM in the Milky Way and in external galaxies can be
made.
\par 
A bright, foreground star was present in the field of NGC~247.  Pixels
contaminated by emission from this star were masked out and
subsequently ignored in the analysis.  H$\alpha$ continuum--subtracted,
mosaiced images of NGC~7793 and NGC~247 are shown in Figure~1 at two
different surface brightness cuts.

\section{Isolating the Diffuse H$\alpha$ Emission}

The separation of total H$\alpha$ flux into that from discrete HII
regions and that from diffuse emission is somewhat complex and subtle
and there is as yet no standard procedure.  Various criteria have
previously been adopted, for example based on the strength of forbidden
line ratios (WB94) or on the equivalent width of the H$\alpha$ line
(Veilleux {\it et al.} 1995).  The  current limited information
available concerning how DIG properties and excitation vary within and
between galaxies prevents a rigorous assessment of how appropriate
either of these methods is for isolating the diffuse emission in a
truly unbiased manner.  A much simpler approach to the separation,
based on surface brightness, is adopted here.  We experimented with our
images until we found a simple isophotal cut which eliminated the bulk
of the discrete HII region population.  This cut was made at a surface
brightness of $1.6 \times
10^{-16}$~erg~s$^{-1}$~cm$^{-2}$/{\sq\arcsec}, uncorrected for [NII] or
extinction, corresponding to an EM of 80~pc~cm$^{-6}$ per pixel (for an
assumed ${\rm T_{e}=10^4}$~K); most of the H$\alpha$ emission lying
below this could be classified as filamentary and/or diffuse.  For
comparison, this isophotal cut is lower than the limits employed by
recent studies of discrete HII region populations in spiral galaxies
(Scowen 1992; Kennicutt {\it et al.} 1988) and also lower than the
isophotal limit which WB94 found isolated the bulk of the diffuse
H$\alpha$ emission in M31 (EM~$< 100$~pc~cm$^{-6}$). Hence, the present
approach to the separation of the two components of the H$\alpha$
emission is a conservative one.  Figure 2 illustrates the diffuse
H$\alpha$ emission in high--resolution subsections of each galaxy, by
masking out pixels having EM $>$ 80~pc cm$^{-6}$.
\par

The moderate inclinations of the galaxies studied here do not allow us
to determine whether diffuse H$\alpha$ emission is also superimposed on
discrete HII regions, though this is very likely to be the case.  We
point out that the observed DIG properties are derived only from pixels
uncontaminated by the emission of HII regions, and hence are strictly
lower bounds to the true values.
\par  
It is of great importance to verify that the diffuse H$\alpha$ emission
we detect is indeed a distinct source of emission (ie. ionized locally)
and not simply recombination photons which are scattered out of
discrete HII regions.  Following WB94, we consider two possible
distinct mechanisms for light scattering that could occur.
\par
First, light could be scattered in the telescope and detector system,
producing extended wings to the point spread function and 
possibly containing a significant fraction of the total light.  To
check for such an effect, we carried out aperture photometry of several
bright stars in the H$\alpha$ frames of both galaxies.  The stars chosen
were isolated and far removed from any galaxian light. In every case,
at least 95\% of the total emitted light was enclosed within an
aperture of radius 11~pixels, corresponding to 4.8\arcsec.    The
results of the aperture photometry were used to derive
azimuthally--averaged radial surface brightness profiles, which
revealed that the total decline in surface brightness ($\sim
10$~mag/\sq\arcsec) occurred over a radius of only 15~--~20~\arcsec.
Beyond this radius, much less than one percent of the total H$\alpha$
light remained. As a result, if the diffuse emission were merely
H$\alpha$ light from HII regions, scattered by the telescope optics,
then it should be highly localised around those regions and the
luminosity of the diffuse component should be significantly less
than that of the discrete HII regions.   As we will discuss below,
there is no evidence for either of these requirements.  Furthermore, the
distinct structure of the brighter diffuse emission -- filaments,
bubbles, loops -- cannot be explained if the diffuse emission were
simply the result of scattered H$\alpha$ photons in the telescope and
detector system.
\par 
The second type of scattering which could occur is reflection from dust
grains in the HII regions themselves.   Since there is no evidence for
halos around bright OB associations in the R--band continuum images,
which should suffer equally, if not more, from the effects of dust
scattering than the H$\alpha$ images, we conclude that the role of this
second type of scattering is also negligible.  We will henceforth refer
to the diffuse H$\alpha$ emission we have detected as diffuse ionized gas
(DIG).

\section{Results}
\subsection{The Morphology of The Diffuse Ionized Gas}

The observed morphology of the DIG, and its variation within and
between galaxies, provides important clues to its origin.  Our deep
H$\alpha$ images reveal a wealth of structure in the ionized gas
components of the galaxies studied here. 
\par
The HII regions in NGC~247 are small, faint and widely separated.  The
brightest HII regions have H$\alpha$ luminosities of $\sim 5 \times
10^{38}$~erg/s, which is slightly lower than the luminosities of the
brightest Galactic HII regions, and considerably lower than the
luminosities of the brightest HII regions found in late--type galaxies
such as the LMC and M101 (Kennicutt 1984).   Much of the DIG in NGC~247
is localized around individual star--forming regions, enveloping them
in frothy, filamentary halos.   The HII regions and the DIG are
largely concentrated into two spiral arms.  In the inner regions of the
disk and in the spiral arms, the DIG halos merge together at faint
intensity levels.  We do not detect any DIG emission from the interarm
regions.
\par
The disk of NGC~7793 is characterized by many luminous HII
regions/complexes (typical luminosities $\ga 10^{39}$~erg/s) and
bright DIG uniformly fills the spaces in between them, out to more than
half of the optical radius of the galaxy.  
This high covering factor of DIG means
that it is difficult to associate a given patch of DIG emission with
any particular star--forming region in the inner regions of the disk.
Further out, the bright DIG tends to be more associated with individual
HII complexes, but all outer disk structures merge together at 
faint levels of a few~pc~cm$^{-6}$.  NGC~7793 lacks well--defined spiral
structure and is instead characterized by a chaotic pattern of
arm fragments.    This is reflected in both the distribution of
HII regions and the distribution of the DIG.
\par
The HII regions in NGC~7793 fill a higher fraction of the disk area
than do those in NGC~247.   As discussed below in Section~5.3, NGC~7793 has
a global star formation rate roughly three times higher than that
inferred for NGC~247.  Furthermore, this star formation is occurring over
a physical area which is less than half the size of that over which
star formation is occurring in NGC~247.   Qualitatively, it appears that
the DIG intensity and covering factor is strongly tied to the star
formation rate per unit area in these galaxies.  
\par
In addition to the relatively unstructured DIG, both galaxies show some
evidence for bubbles, loops  and filamentary features, extending up to
a few hundred parsecs in size.   Such features are considered clear
evidence for the action of massive stars on the environment
(Tenorio--Tagle and Bodenheimer 1988).   Some of these features have
sufficiently high surface brightnesses that they lie above our surface
brightness limit for the DIG and hence their emission is classified as
being due to discrete HII regions. The features we observe  in NGC~7793
and NGC~247 are much less spectacular than features we have identified
in two other Sculptor disk galaxies, NGC~55 and NGC~300 (Ferguson et
al., in preparation).  They are also smaller in size than the
supergiant shells which have been identified in the LMC, a galaxy which
is of similar absolute magnitude and present star formation rate to the
ones studied here, but lacking in spiral arms (however several of these
supergiant shells have surface brightnesses which lie above our
criterion for diffuse emission (Hunter 1994)).  It is tempting to
speculate that the sizes of the filamentary and bubble--like features,
which constitute the structured component of the DIG, are a direct
result of the intensity of local star formation, however magnetic
fields may play an equally important role in regulating their
morphologies (Elmegreen 1987; Hunter and Gallagher 1990).

\subsection{The Distribution of Diffuse Ionized Gas}
\subsubsection{Azimuthal Variation}

The azimuthal variation of H$\alpha$ emission provides important
information on the uniformity of the intensity distribution of the DIG
and on the spatial association between bright HII regions and the DIG.
Figure~3 shows the azimuthal variation of H$\alpha$+[NII] emission
measure around a series of four elliptical annuli, chosen to span a
large fraction of the optical disk in each galaxy.  The emission
measures have not been corrected for extinction.  The radial thickness
of each annulus is 250~pc and the data are binned in 2{$\deg$}  sectors
around the azimuthal direction.   The adoption of a fixed angle for
azimuthal binning results in a variable smoothing factor in terms of
pixels.  For example, in NGC~7793, the azimuthal bin size corresponds
to physical areas of size $\sim 250$~pc $\times$~15~pc and
$\times$~135~pc in the annuli centered on 0.1R$_{25}$ and  0.9R$_{25}$
respectively.  In NGC~247, the bin sizes are $\sim$~250~pc
$\times$~19~pc and $\times$~175~pc in the 0.1R$_{25}$ and 0.9R$_{25}$
annuli.      For reference, small Galactic HII regions have typical
diameters of a few tens of parsecs and hence would be smoothed over in
our analysis.  While this may account for some of the low surface
brightness emission seen in the azimuthal profiles in Figure 3, visual
inspection of the images reassures us that the bulk of this emission
arises from truly diffuse and filamentary structures.  The largest HII
region complexes found in late--type spirals have diameters of order a
few hundred parsecs (Kennicutt 1984) and hence would be preserved in
our azimuthal plots, if centered.

\begin{figure}[h]
\figurenum{3}
\plottwo{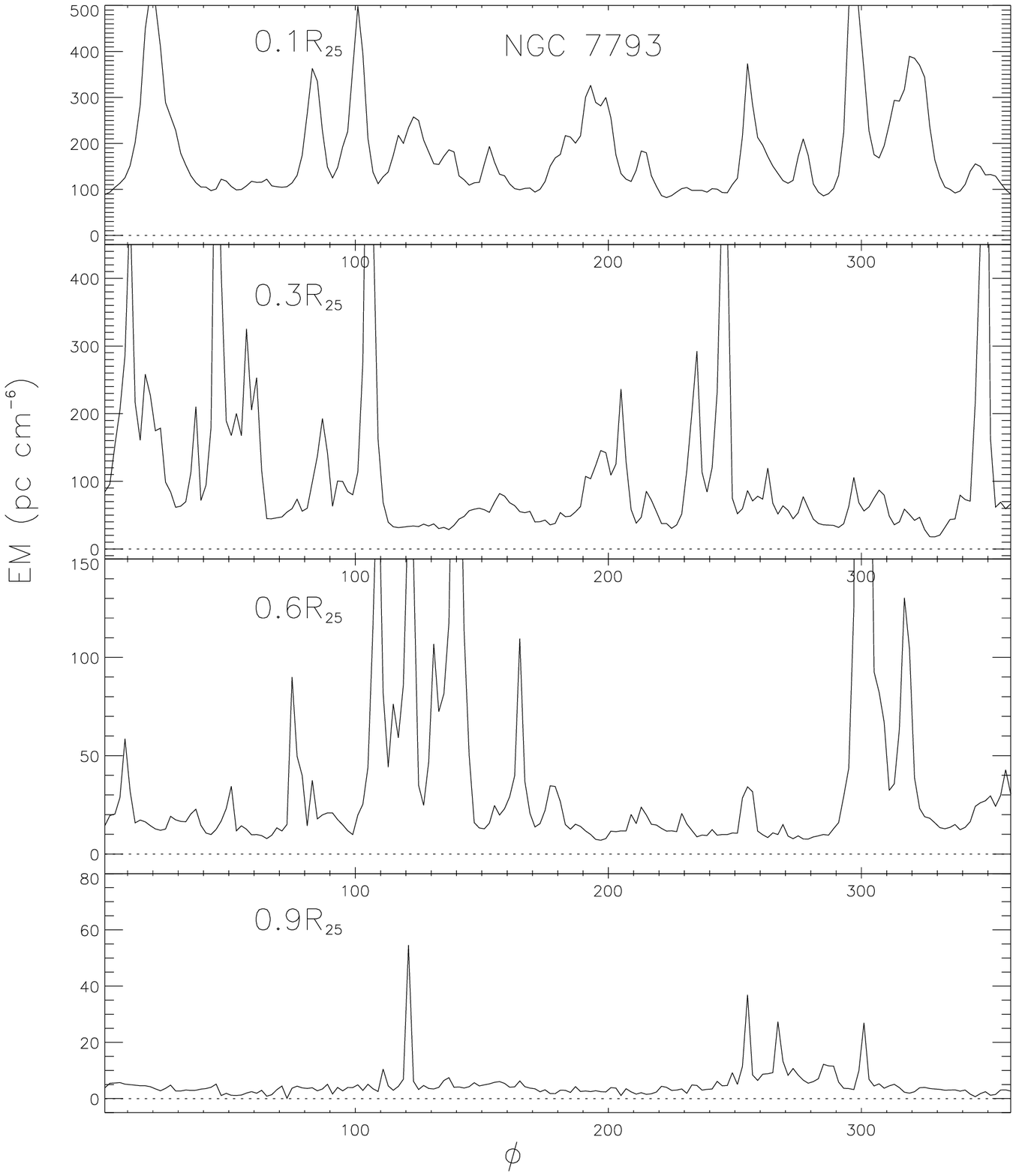}{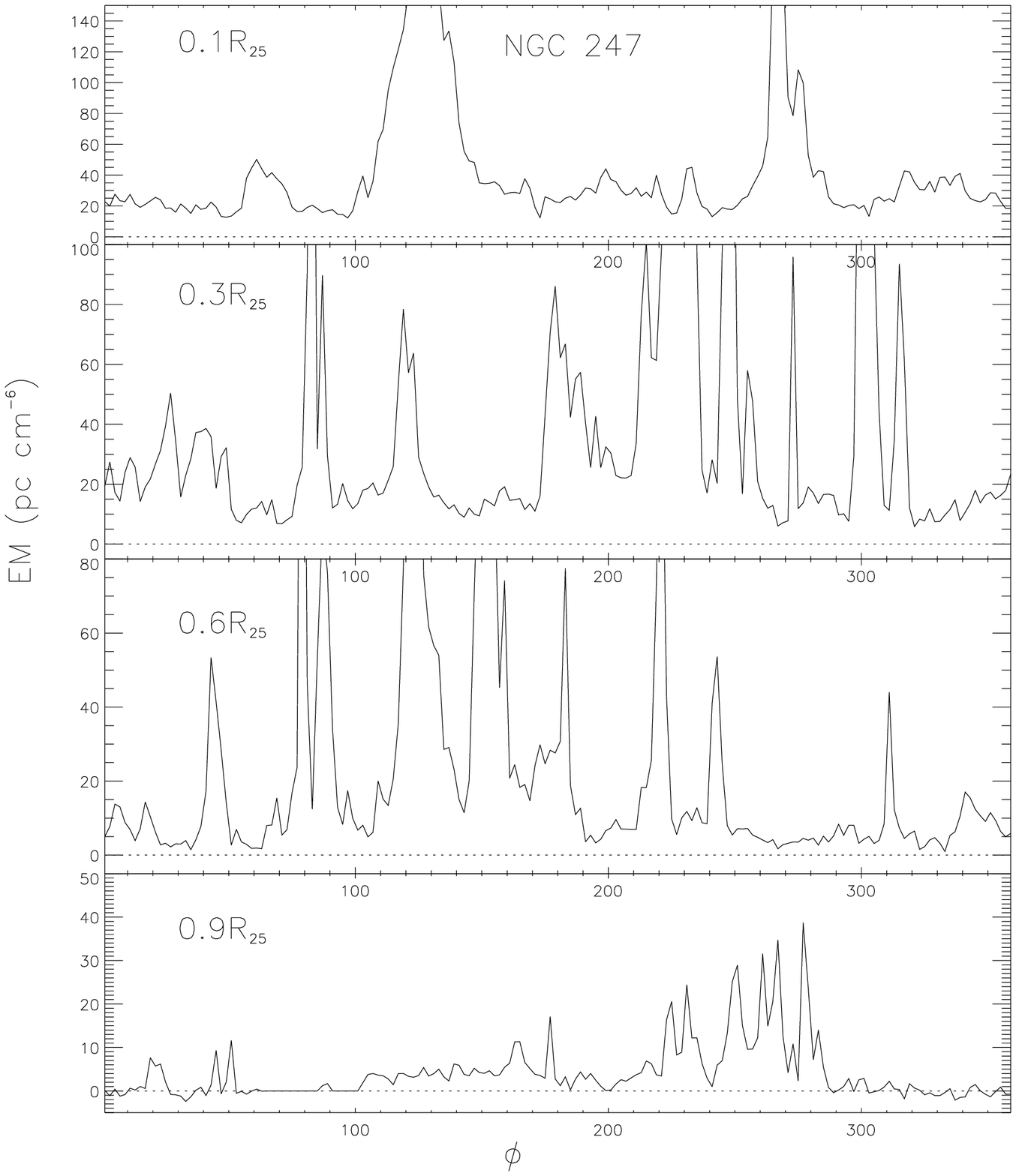}
\caption{(a) Plot of the azimuthal variation of uncorrected
H$\alpha$ +[NII] surface brightness at four radii in NGC~7793 .  The
radii correspond to 0.1R$_{25}$, 0.3R$_{25}$, 0.6R$_{25}$ and
0.9R$_{25}$.  The width of each radial bin is 250~pc and the data are
azimuthally binned into 2{$\deg$} sectors.  (b)  As in (a) but for NGC
247.}
\end{figure}
 
Two key results emerge from the azimuthal analysis.  Firstly, in the
inner radial bins of both galaxies, the DIG fills the annuli uniformly
defining a {\it baseline} in the value of the emission measure. This
baseline is $\sim$~100~pc~cm$^{-6}$ in NGC~7993 and
$\sim$~20~pc~cm$^{-6}$ in NGC~247.  As the radial distance from the
center of each galaxy increases, this baseline value gradually
decreases and in the outermost bins here, the H$\alpha$ emission often
becomes too faint to detect far from bright HII regions and complexes.
Secondly, our azimuthal plots reveal that bright HII regions/complexes
often have substantial halos which blend smoothly into the DIG.  The
brightest DIG emission appears to be closely associated with
sites of recent star formation and the intensity of the DIG is observed to
fall off with distance from the nearest HII region.  The plots confirm
that, in the inner disks of both galaxies, it is impossible to
associate a given patch of DIG emission with any particular HII
region.  At larger galactocentric distances, however, much of the DIG
is localized around bright OB complexes, although these are few in
number.
\par   
The radial and azimuthal smoothing above, while facilitating
qualitative analysis, hampers the derivation of quantitative
information on the typical scales on which DIG is associated with
individual HII regions.  Direct inspection of the images (Figures 1 and
2) reveals that DIG can be found up to distances in the range of 200 --
900~pc from bright HII regions, with a typical extent of around 500~pc.
Clearly, the brighter DIG is associated with HII regions, but at
fainter levels, it becomes ubiquitous.

{\subsubsection{Radial Variation}}

In order to study the radial variation of the  ionized gas components
in these galaxies, we carried out elliptical aperture photometry on our
H$\alpha$ images,  using ellipses fit to the corresponding R--band
images. This approach was chosen since the H$\alpha$ images themselves
are too clumpy for meaningful ellipses to be fit. The radial step size
was 350~pc for NGC~7793 and 500~pc for NGC~247.   Radial surface
brightness profiles were derived for the total H$\alpha$ emission, the
H$\alpha$ emission from HII regions alone, and the H$\alpha$ emission
from the DIG (using the isophotal surface brightness criterion
discussed previously).  In calculating the DIG surface brightness, we
normalised to the number of pixels that had a surface brightness less
than 80 pc cm$^{-6}$, as opposed to the total number of pixels per
annulus.  This approach freed us from making assumptions about the
intensity of the DIG that is superimposed on  bright HII regions.  A
deprojection was carried out (cos{\it(i)} correction) using the values
of the inclination listed in Table 1.  We note that if the DIG emission
arises in a thick--disk component then the thin-disk correction we have
applied will result in an underestimate of the true radial surface
brightness amplitude.  In this case, the deprojection becomes a
complicated function of the radial and vertical profiles of the ionized
gas disk and the dust.
\par
The discrete and diffuse H$\alpha$ emission was corrected separately
for contamination by the [NII] lines and for the effects of Galactic
and internal extinction.  Webster and Smith (1983) have presented
spectra for various HII regions in the disk of NGC~7793.  We have used
these data to derive the disk-averaged I([NII])/I(H$\alpha$) ratio and
the Balmer decrement, C$_{H\beta}$, for the HII regions in NGC~7793
with the result that [NII]/H$\alpha_{HII}$ = 0.22 and
C$_{H\beta}$=0.38.   We note that the data available indicate no
significant change in the [NII]/H$\alpha$ ratio across the disk (even
although the galaxy has a moderately steep metallicity gradient).   The
reddening law of Schild (1977) then gives an extinction of
A$_{H\alpha}$=0.64.   Spectroscopic information does not exist for any
HII regions in NGC~247, so the {NII]/H$\alpha$ ratio and the mean
extinction values derived for NGC~7793 were assumed equally applicable
for NGC~247.  Thus, the derived quantities for NGC~247 are more
uncertain than those for NGC~7793.  There has not been a direct
measurement of the [NII]/H$\alpha$ ratio in the DIG component of the
galaxies presented here; however, studies of other galaxies indicate
that this ratio is often systematically higher than that measured for
discrete HII regions (Dettmar 1992; Dettmar and Schultz 1992; Veilleux
{\it et al.} 1995;  Hunter 1994).  We adopt [NII]/H$\alpha_{DIG}$=0.50
to be consistent with the mean values found in those studies.  In the
absence of data which can be used to constrain directly the extinction
of the diffuse gas (eg. WB94), we adopt the same extinction correction
for both the DIG and the discrete HII regions unless otherwise noted.
Figure 4 shows the radial variation of the ionized components measured
out to the optical radius and corrected for our adopted [NII]
contamination and extinction.   In both galaxies, we clearly detect DIG
over a large fraction of the star-forming disks (ie.  to
$\sim$~R$_{25}$).

\begin{figure}[h]
\figurenum{4}
\plottwo{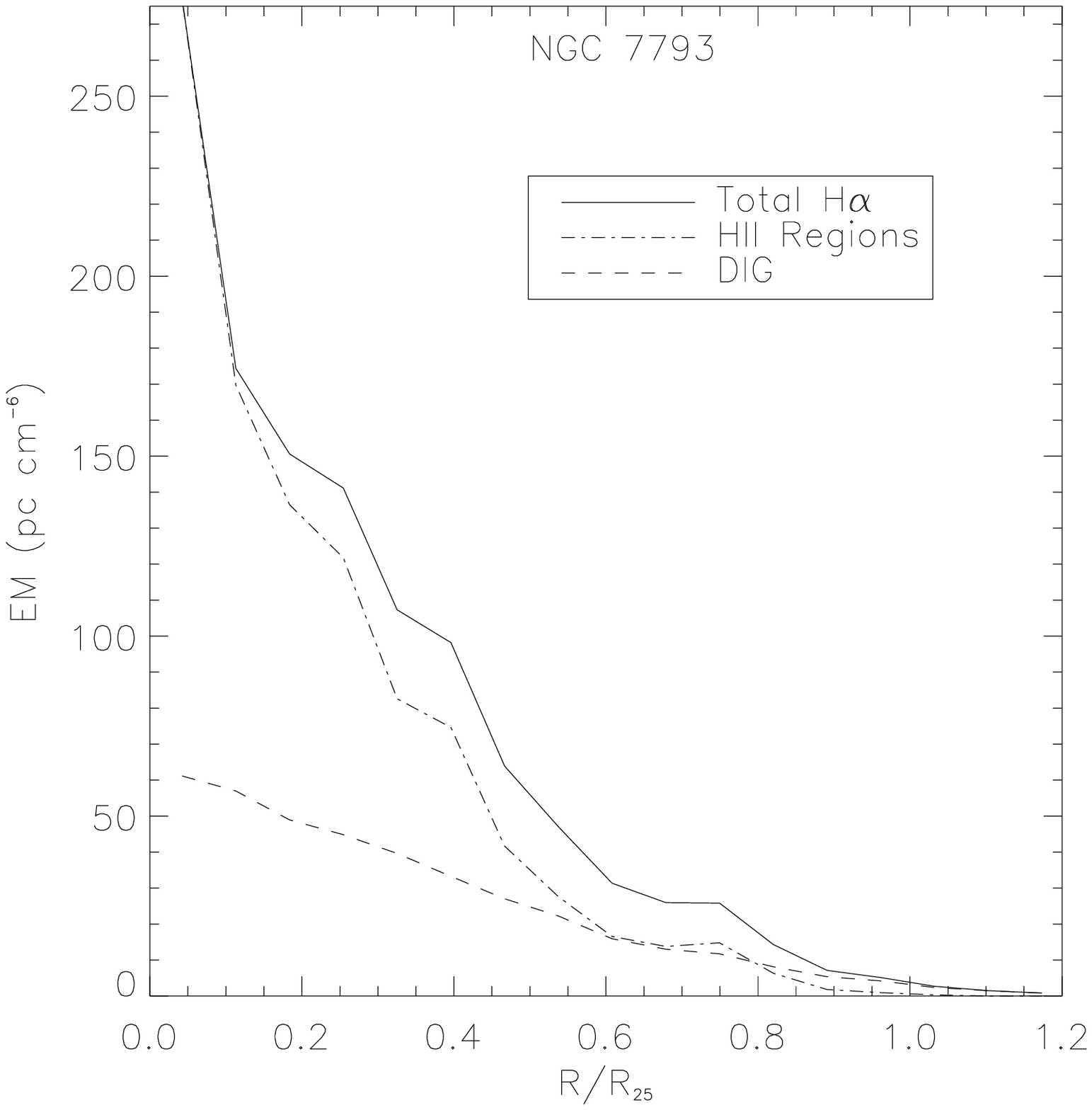}{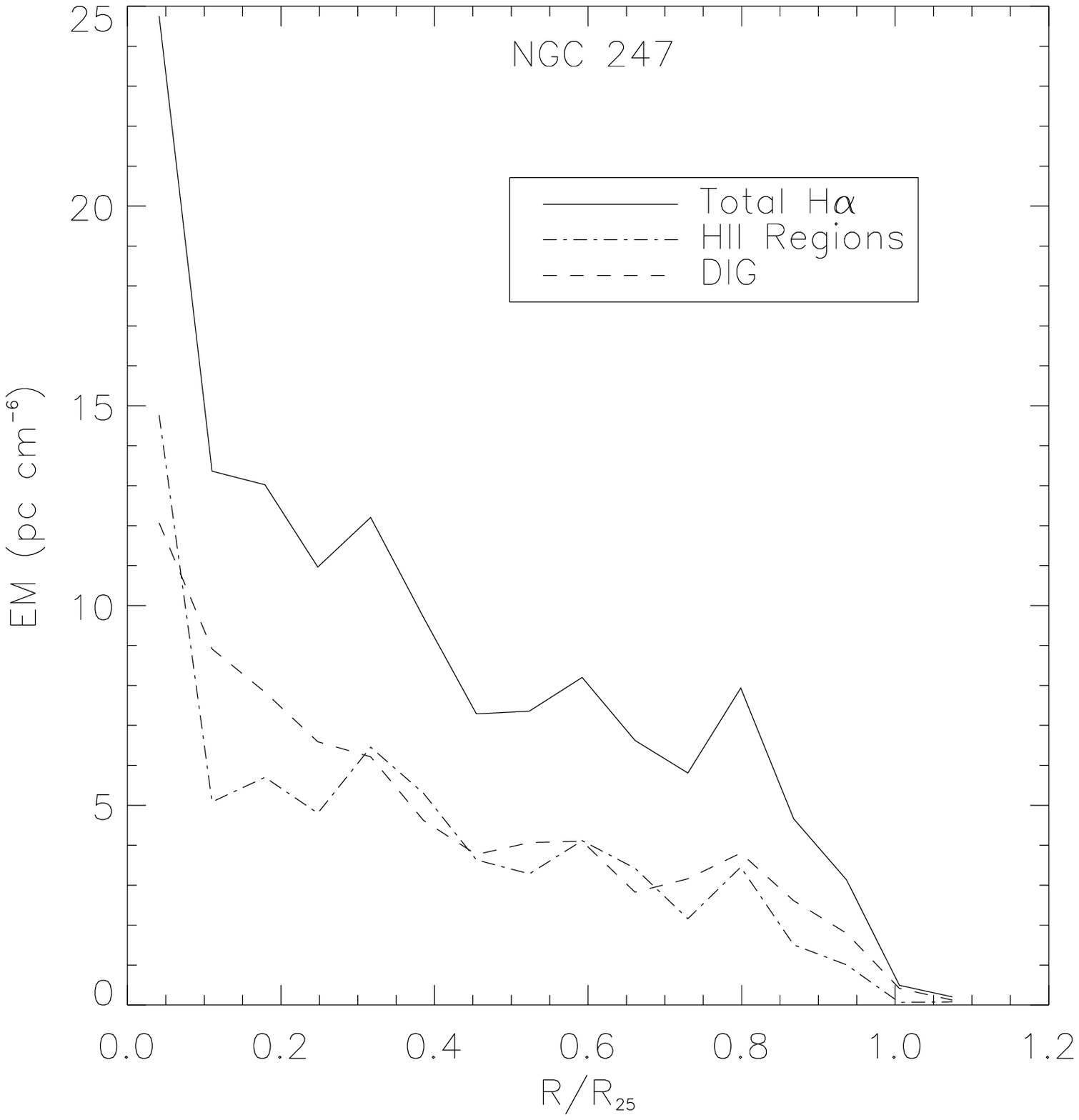}
\caption{(a) Deprojected radial profile of the total
H$\alpha$ surface brightness (solid line), the H$\alpha$ surface
brightness derived by counting HII regions alone (dashed--dotted line)
and the H$\alpha$ surface brightness of the DIG (dashed line) in NGC
7793.  The optical radius (R$_{25}$) is indicated.  (b)  As in (a) but for
NGC~247.}
\end{figure}

Very different H$\alpha$ profiles are observed across the two
galaxies.  NGC7793 has a centrally peaked H$\alpha$ profile with a
region of very active star formation, as traced by the HII regions,
occurring out to a galactocentric radius of $\sim$~3~kpc.  The total
average emission measure falls off quite steeply, ranging from
$\gtrsim$ 250~pc cm$^{-6}$ in the center of the galaxy, to a few~pc
cm$^{-6}$ in the outer parts of the disk.  The DIG emission measure
ranges from  $\sim$ 60~pc~cm$^{-6}$ in the center, to a few~pc
cm$^{-6}$ in the outer galaxy.  The DIG profile in NGC7793 is much
flatter than that of the HII regions in the inner disk, but the
profiles become very similar beyond a radius of $\sim$~2.5~kpc.    Part
of the central flattening must be due to the present definition of DIG;
some high surface brightness  (EM $\simeq$~150~pc cm$^{-6}$), truly
diffuse emission is clearly visible in the inner regions of the galaxy
but has been misclassified by our separation technique. If properly
included, the two inner profiles would possibly show a more similar
behaviour.
\par
On the other hand, NGC247 is characterized by very low H$\alpha$
surface brightnesses across its disk, at least a factor of 10 below
those observed in NGC7793 at similar radii, and has an extremely flat
profile with only a weak central enhancement.   The radial distribution
of the diffuse ionized component closely follows that of the HII
regions throughout the entire disk.  The DIG emission measure ranges
from  12~pc~cm$^{-6}$ in the center to a few~pc~cm$^{-6}$ in the outer
regions, with the bulk of the disk having a DIG surface brightness  of
around 5~pc~cm$^{-6}$.  Thus, in NGC~247 we have made a detection of an
extragalactic ionized hydrogen disk of comparable surface brightness
to that measured locally in the Reynolds layer ($\sim$~6~pc~cm$^{-6}$;
Reynolds 1984).
 
\subsection{The Diffuse Fraction}

\subsubsection{Global Values}

The observed total luminosity of the DIG in each galaxy was obtained by
summing the intensities of pixels with observed surface brightnesses
corresponding to an EM $<$ 80~pc~cm$^{-6}$ and then correcting for
contamination by [NII] and extinction as described above.  The observed
H$\alpha$ DIG luminosities are found to be $7.5 \times 10^{39}$~erg~s
$^{-1}$ for NGC~247 and $1.2 \times 10^{40}$~erg~s$^{-1}$ for NGC~
7793. The observed global diffuse fractions, defined to be the 
fraction of the total H$\alpha$ luminosity contributed by the DIG, are
50\% for NGC~247 and 29\% for NGC~7793.   Lower limits on the 
diffuse fraction may be derived by the assumption of zero extinction
for the diffuse gas and are 37\% for NGC~247 and 19\% for NGC~7793.
\par
As previously mentioned, the observed DIG H$\alpha$ luminosities are
strictly lower bounds to the true H$\alpha$ luminosities since they are
derived on the assumption that the intensity of the DIG is zero on top
of discrete HII regions.  A more realistic estimate of the luminosity
of the DIG can be derived as follows.   We assume that each
elliptical annulus is filled with DIG at the mean DIG surface
brightness measured at that particular radius (as indicated in the DIG
surface brightness profiles presented in Figures 4a and b).   In
essence, this amounts to counting a fraction of the flux from each HII
as being emission due to superimposed DIG.  This is a small effect everywhere
except in the innermost bins, where the HII region covering--factor is
significant, and is still likely to give an underestimate of the true
H$\alpha$ luminosity of the DIG.  The total DIG luminosities of all
the individual annuli are then summed and corrected for [NII] and
extinction.  The resulting {\it corrected} DIG H$\alpha$ luminosities
are  $7.9 \times 10^{39}$~erg~s$^{-1}$ for NGC~247 and $1.7 \times
10^{40}$~erg~s$^{-1}$ for NGC~7793.  We note that the corrected DIG
H$\alpha$ luminosity in NGC~247 is only slightly larger than the
observed one; this is due to the small covering factor of bright HII
regions across the disk of the galaxy.   These values can be used to
calculate the corrected global diffuse fractions with the result of
53\% for NGC~247 and 41\% for NGC~7793.
\par
The global diffuse fractions derived here are very similar to the
global diffuse fractions found in other actively star--forming nearby
galaxies (but being mindful of the wide range of assumptions and
corrections that different authors have applied):  40\% in
M31 (WB94); 30\% in NGC~3079 (Veilleux {\it et al.} 1995); 35\% in the
LMC;  41\% in the SMC (Kennicutt {\it et al.} 1995) and  35\%  and 50\%
in the Magellanic Irregulars NGC~4214 and NGC~4449 (Kennicutt {\it et
al.} 1989). (Note that Hunter and Gallagher (1990) find lower diffuse
fractions, 15\% -- 20\%, in NGC~4214 and NGC~4449; this may be due to
the lower isophotal surface-brightness cut they used to define the
DIG.)  Diffuse fractions in this general range were also inferred for a large
sample of nearby spirals on the basis of emission--line ratios in their 
integrated spectra (Lehnert \& Heckman 1994).  The similarity in the
measured diffuse fractions is somewhat surprising given the wide range of
morphological types (Sb -- Im) and star formation rates (0.01 -- 0.6
M$_{\sun}$ yr$^{-1}$) spanned by these galaxies.  Furthermore, several
of these galaxies have large~kpc-scale ionized bubble and filamentary
features (NGC~3079, LMC, SMC, NGC~4449) while others do not.  Such
obvious signs of HII region disruption do not appear to lead to
relatively larger quantities of DIG.  One possible interpretation of
this result is that the overall fraction of photons which leak out from
HII regions is relatively constant from galaxy to galaxy, appearing at
first glance to be largely independent of star formation and ISM
properties.  Alternatively, it could reflect the fact that these
galaxies all have a significant fraction of their OB star population
residing outside the boundaries of classical HII regions (eg. Patel and
Wilson 1995).

\subsubsection{Radial Variation}

The variation of the diffuse fraction within a given galaxy is of
considerable interest since it constrains the origin of the DIG.  We
calculated this quantity by taking the ratio of DIG luminosity to total
H$\alpha$ luminosity in the series of elliptical annuli used for
deriving the radial profiles.  In view of the artificially low values
of the diffuse fraction we derived in the inner regions, where few
pixels are uncontaminated by bright HII regions, we assumed again that
the DIG emission filled each ellipse at the mean surface brightness
measured.    Figure
5 shows the radial variation of the diffuse fraction across the disks
of NGC~7793 and NGC~247.   The diffuse fraction is almost constant at
$\sim$~50\%  across the disk of NGC~247, whereas  it rises in NGC~7793
from $\sim$~30\% in the inner regions to $\sim$ 90\% in the outermost
parts.  The suspected under--counting of DIG due to misclassification
in the inner disks may account for part of the radial trend seen in
NGC~7793 but cannot explain the entire amplitude.

\begin{figure}[h]
\figurenum{5}
\plotfiddle{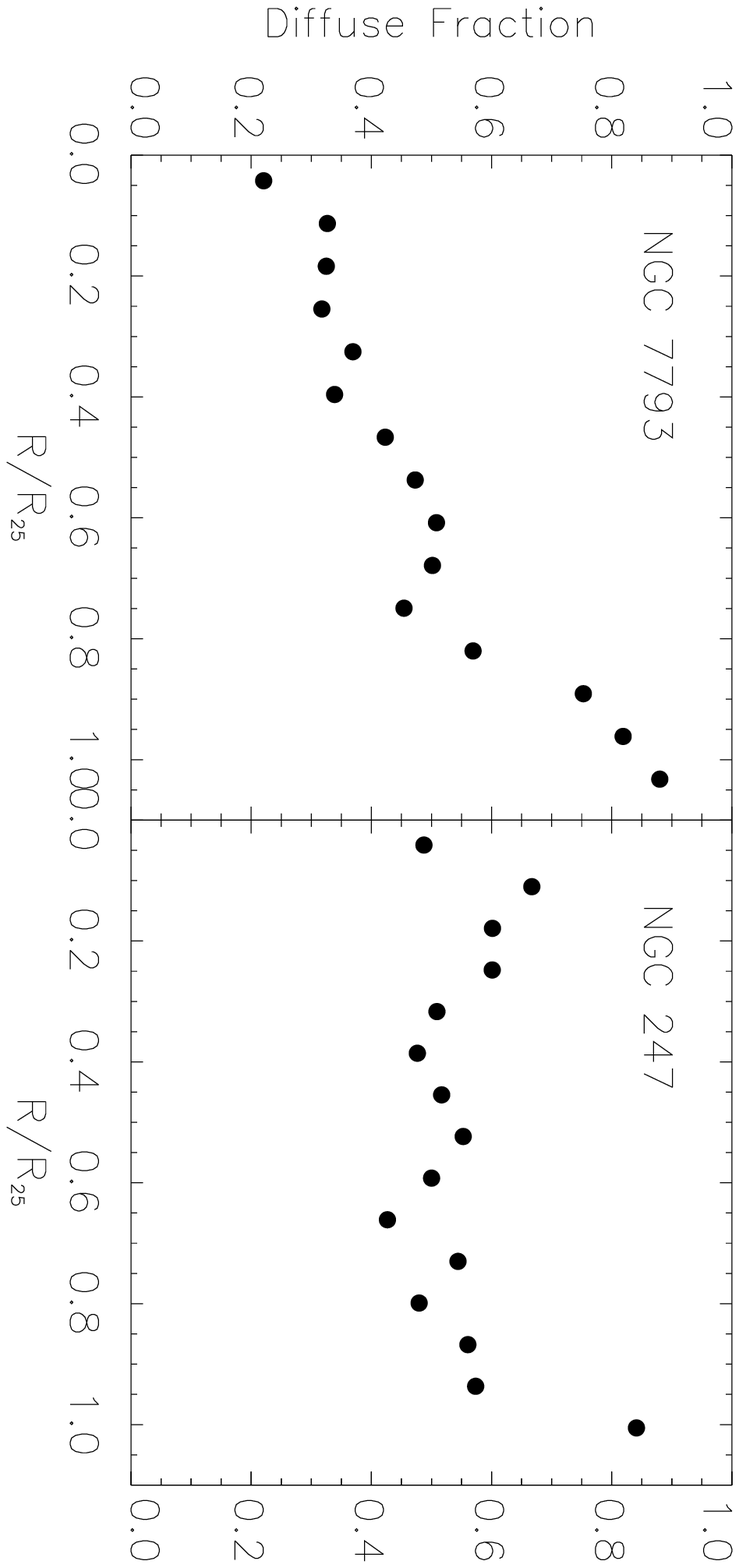}{7cm}{90}{65}{65}{255}{0}
\caption{Radial variation of the diffuse fraction for NGC~7793
(left) and NGC~247 (right).}
\end{figure}

Our results for the radial variation of the diffuse fraction are in
qualitative agreement with the predictions of recent models for the
photoionization of the DIG by Lyc photon leakage from OB associations
(Dove and Shull, 1994).  These models incorporate a smoothly varying HI
distribution, and thus are limited in the sense that they do not allow
for structure in the HI or for a population of opaque clouds.  In these
models, the radial profile of HI plays a role in determining the radial
variation of the diffuse fraction, with the fraction of `escaped' Lyc
photons being essentially proportional to a constant plus R/R$_{g}$,
where R$_{g}$ is the HI gas scale--length (their equation 30).  If the
HI profile is flat, ie.  R$_{g}$=$\infty$, then the `escaped' fraction
is predicted to be constant across a disk.  Both galaxies discussed here have
declining HI profiles over their optical disks.  The decline in
NGC~7793 is much steeper (ie.  shorter scale--length) and takes place
over a larger fraction of the disk than that observed in NGC~247
(Carignan \& Puche 1991; Puche \& Carignan 1991).  As a result, the
Dove and Shull (1994) models predict that the diffuse fraction should
rise more steeply in NGC~7793 than in NGC~247, consistent with the
observations.

\subsubsection{Errors}

It is important to quantify the effects that small errors in the
flat--fielding, sky subtraction and/or the continuum subtraction could
have on the derived luminosities and hence the global diffuse fractions
presented above.    

As discussed previously, the sky value is determined for each combined
H$\alpha$ and continuum frame, and subtracted off before the continuum
subtraction and mosaicing is carried out.  Any small error in the sky
determination for a given frame could become more significant in the
final H$\alpha$ continuum--subtracted mosaiced image.  As discussed in
Section 2,  we determined the sky level across the final mosaiced
images using small (100 $\times$ 100 pixels) boxes situated beyond the
Holmberg diameters of each galaxy.  The sky values obtained were
0~$\pm$~1.5~pc~cm$^{-6}$ in the NGC~7793 frame and
0~$\pm$~3.2~pc~cm$^{-6}$ in the NGC~247 frame.   Assuming a maximum
error in the sky of $\pm$~1.5~pc~cm$^{-6}$ across the entire field of
NGC~7793,  the resulting error in the corrected DIG H$\alpha$
luminosity would be $\sim$~6\% , the error in the total H$\alpha$
luminosity would be $\sim$~2\%  and the error in the corrected global
diffuse fraction would be $\sim$~2\%.  Likewise,  an assumed maximum
error of $\pm$~3.2~pc~cm$^{-6}$ across the entire field of NGC~247
leads to an error in corrected DIG H$\alpha$ luminosity of $\sim$~32\%,
an error in total H$\alpha$ luminosity of $\sim$~17\%  and an error in
the corrected global diffuse fraction of $\sim$~8\%.    The larger
errors for derived quantities in NGC~247 result from a combination of a
larger sky error per pixel and a larger number of pixels covered by the
galaxy.   While these simple calculations neglect the fact that
adjusting the sky value also affects our classification of emission as
being either diffuse or discrete, they serve to give a rough idea of
the stability of our derived quantities to realistic errors in the sky
background.
\par
Another source of possible uncertainty lies in the continuum scaling
factor used to produce the final H$\alpha$ continuum--subtracted
images.  As previously discussed, the continuum scaling factor was
derived by measuring the ratio of the fluxes of foreground stars in the
H$\alpha$ and continuum images.  The scale factor was slightly adjusted
until the bulk of the foreground stars were perfectly subtracted and
there were no large positive or negative regions across the galaxy, as
judged by eye.  We experimented with changing the scaling factor by
$\pm$~2\% and then by 7\%, and noted how this affected our derived
quantities.  For both galaxies,  a 2\% change in the scaling factor led
to changes of $\sim$~5--8\% in the corrected DIG H$\alpha$
luminosities, $\sim$~5\% in the total H$\alpha$ luminosities and no
appreciable change in the corrected global diffuse fractions.  Changing
the continuum scaling factor by 7\% led to changes of $\sim$~25\% in
the corrected DIG H$\alpha$ luminosities, $\sim$~15\% in the total
H$\alpha$ luminosities and $\sim$~5\% in the corrected global diffuse
fractions.  We note that a 7\% change in the continuum scaling factor
corresponds to the point at which spurious large scale positive and
negative variations become clearly visible across the galaxies and
foreground stars are improperly subtracted; thus, this is the maximum
possible error on the derived continuum scaling factor.    Application
of a single continuum scaling factor across the disk of the galaxy, as
done here, implicitly assumes no gradient in the H$\alpha$ equivalent
width.  This may not be entirely appropriate but is expected to be a
low amplitude effect.
 
\subsection{Lyc Power Requirements}

The deprojected, [NII]-- and extinction-corrected radial profiles of the
DIG in these galaxies can be used to calculate the {\it minimum} power
per unit area required to keep the DIG ionized.  The recombination rate
per cm$^2$ of disk, $r$,
is first calculated using the relation 
\begin{displaymath}
r~=~{{4~\pi}\over{\epsilon}}~{I_{H\alpha}},
\end{displaymath}
where $\epsilon$ is the average number of H$\alpha$ photons per
recombination (assumed to be $\sim$~0.46; Case B recombination at
10$^{4}$~K) and I$_{H\alpha}$ is the H$\alpha$ surface brightness in
units of photons~cm$^{-2}$~s$^{-1}$~sr$^{-1}~$; the power consumption
per unit area is then derived under the assumption that every
ionization requires a minumum input energy of 13.6~eV.  These power
requirements are plotted in Figure 6, where the dashed line indicates
the value estimated for the solar neighborhood, $1 \times
10^{-4}$~erg~s$^{-1}$~cm$^{-2}$ (Reynolds 1984).   As can be seen,
NGC~7793 has minimum power per unit area requirements which are
significantly higher than the local Galactic value, whereas NGC~247
has requirements which are similar to the Milky Way.

\begin{figure}[h]
\figurenum{6}
\plotfiddle{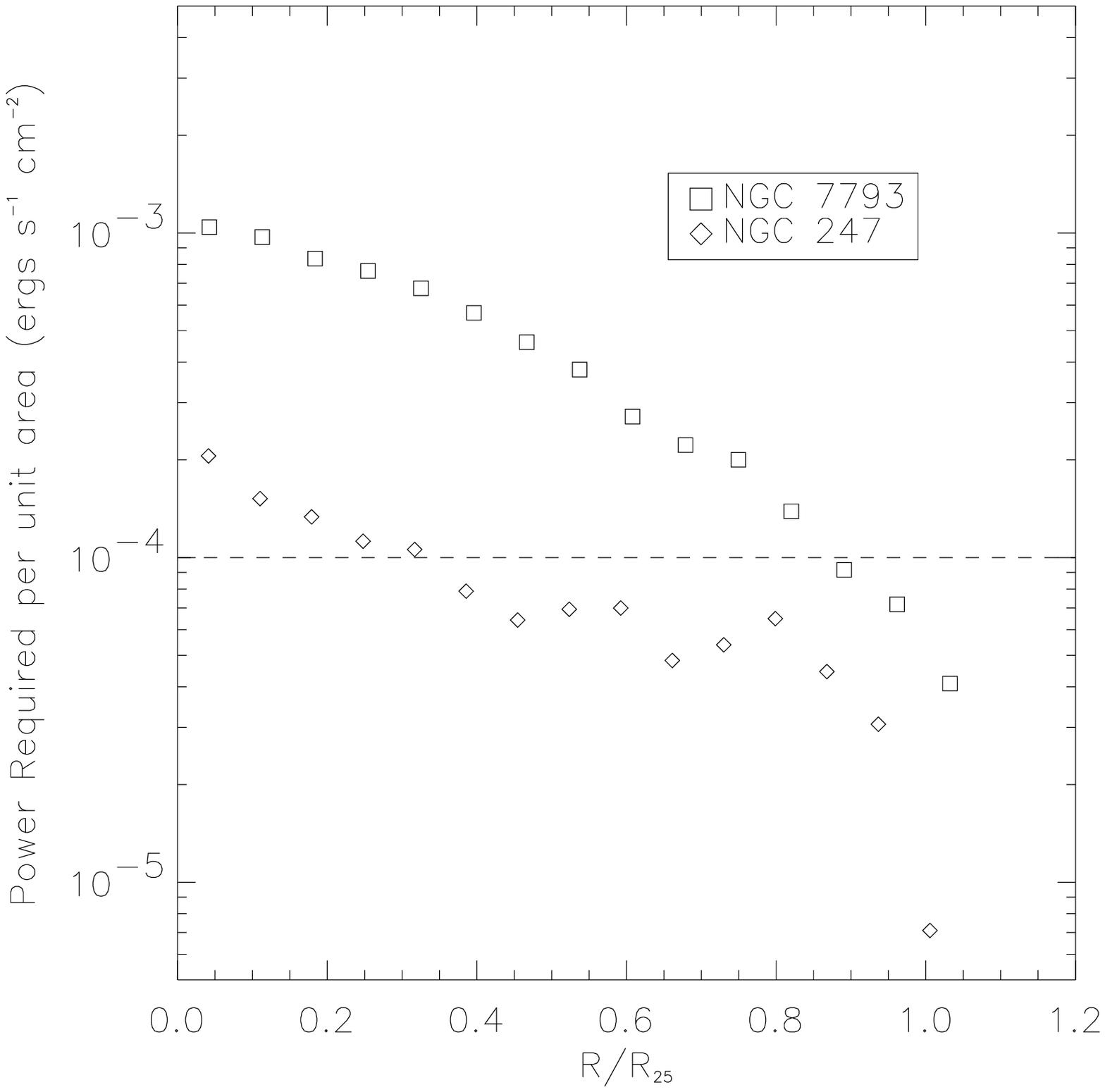}{8cm}{0}{50}{50}{-180}{-55}
\caption{Minimum power per unit area required to maintain
the DIG across the disk of each galaxy.  The dotted line indicates the
value derived for the solar neighborhood (Reynolds 1984).}
\end{figure}

Based on the {\it observed} total DIG H$\alpha$ luminosities discussed in
Section 4.3.1, we can derive the minimum integrated Lyman continuum
power required to keep the DIG ionized, using the same procedure as
outlined above.   We find minimum Lyc power requirements of $1.2 \times
10^{41}$~erg~s$^{-1}$ for NGC~247 and $1.9 \times 10^{41}$~erg~s$^{-1}$
for NGC~7793.  If we use the more realistic {\it corrected} H$\alpha$
luminosities of the DIG,  we derive power requirements of $1.25 \times
10^{41}$~erg~s$^{-1}$ for NGC~247 and $2.7 \times 10^{41}$~erg~s$^{-1}$
for NGC~7793.  These values are all slightly lower than the Galactic
one of $6 \times 10^{41}$~erg~s$^{-1}$ (Reynolds 1984), which was
derived on the basis of extrapolating {\it local} values over the
entire disk.  The galaxies studied here are smaller than the Milky Way
and hence lower global power requirements are not unexpected.
Nevertheless, the enormous amount of power required to sustain the DIG
in these galaxies (at least $3 - 5 \times 10^{7}$~L$_{\sun}$ in Lyc
photons alone) places severe constraints on the possible origin of the
ionizing photons.

\section{Discussion}

\subsection{Small and Large Scale Distribution of the DIG}

The azimuthal and radial intensity distributions of H$\alpha$ emission
in these galaxies indicate that the DIG is highly correlated with
bright HII regions over a variety of scales.  On scales of a few
hundred parsecs, the bright DIG is highly concentrated around
luminous HII regions, with the mean surface brightness decreasing as
the distance to the nearest HII region increases.  This close small-scale
association between DIG and HII regions was also observed by WB94 in
selected fields in M31 and lends strong support for a model in which the Lyc
photons which ionize the DIG leak out of individual HII regions.  In this
picture, the bulk of the H$\alpha$ recombinations would be expected to take
place in the immediate vicinity of the leaking HII region, where the Lyc
photon flux is higher.   At faint levels, the DIG is ubiquitous, though
generally follows the overall pattern of star formation e.g. being enhanced
in spiral arm structures if present, as in NGC~247.
\par
Over larger scales, the DIG has spatial and intensity distributions
which closely follow those of the discrete HII regions in each galaxy.  A key
observation is that the DIG is detected over the entire extent of the bright
star--forming disks in both galaxies.   The faint, outer disk DIG we
detect has surface brightnesses comparable to the large--scale
flat--fielding errors, hence we are unable to address the important
issue of whether the region of DIG emission actually extends beyond the
edge of the HII region distribution.  Deeper CCD data we have just obtained
for these galaxies will allow a more detailed future study of the outer disk DIG
emission. 
\par
The large
scale radial distributions reveal that the DIG intensity falls off with
the mean surface brightness of HII regions and, hence, the mean level of
star formation.   Hester and Kulkarni (1990) and Veilleux {\it et al.} (1995)
have also noted a large scale radial variation of DIG intensity, in
studies of M33 and NGC~3079 respectively, but they did not make a
comparison with the distribution of discrete HII regions in these
galaxies.  The similarity between the DIG and the discrete HII region surface brightness profiles observed further supports a leakage origin for the
ionizing photons which create the DIG.  Moreover, the correlation between
the mean surface brightness of discrete HII regions and the
DIG suggests that large--scale radial photon transport, on scales
comparable to the widths of our annular bins, 250~pc, does not involve
a significant number of ionizing photons.
\par
The small and large scale distributions of the DIG are easily
understood in terms of a model where Lyc photon leakage from
star--forming regions provides the dominant source of the ionizing
photons.  The available evidence indicates that the star formation rate
per unit area (traced by the surface brightness of discrete HII
regions) is correlated with the DIG surface brightness, in the sense
that a higher mean star formation rate per unit area leads to a higher
number of H$\alpha$ recombinations occurring outside of discrete HII
regions.  The leakage model is not the only model which can explain our
observations of the large scale distribution of the DIG,  however, and
any method of producing ionizing photons which is tied to Population I
objects is equally plausible.  A model in which  a significant fraction
of the OB star population resides outside the boundaries of discrete
HII regions may also suffice (eg. Patel and Wilson 1995), but it
remains unclear whether ionization by {\it in situ} OB stars can
explain other DIG properties, such as the observed line ratios.
Decaying neutrinos might be expected to produce an ionized gas
component with a radial profile similar to the density profile of the
dark matter halo, ie.  volume density proportional to the inverse
square of the radius (Sciama \&  Salucci 1990).  While such a
distribution is not necessarily inconsistent with the profiles  derived
here, it is difficult to understand  why the DIG  emission would be
enhanced around individual HII regions, unless the gas becomes
optically thin in these regions as well.

\subsection{The Diffuse Fraction}

The global diffuse fractions of the galaxies presented here, together
 with those found in other galaxies, show a striking similarity to each
other.  If the amount of Lyc photons produced in HII regions were the
only governing factor in determining the properties of the DIG, then
galaxies with low global star formation rates, such as the SMC, would
be expected have different DIG properties than more actively star--
forming galaxies, such as NGC~7793 presented here.   The constancy  of
the diffuse fraction among galaxies of differing Hubble types and star
formation rates suggests a conspiracy between the level of star
formation and some additional factor, with the result that the fraction
of photons which escape from discrete HII regions is more or less
constant from galaxy to galaxy.
\par
A similar conclusion can also be reached from inspection of the radial
variation of the diffuse fractions in NGC~247 and NGC~7793.  Since the
surface brightness of H$\alpha$ emission from HII regions, hence Lyc
production, decreases with increasing galactocentric radius, one might
expect less diffuse emission to be present in the outermost parts.
Instead, the diffuse fraction in NGC~247 is constant across the disk, 
while the diffuse fraction in NGC~7793 increases with galactocentric
radius.  WB94 also found a relatively constant diffuse fraction in each
of the fields they surveyed in M31.  This behavior suggests that the
additional factor which regulates Lyc photon leakage in galaxies is one
which has some radial dependence within disks, and one which varies
from galaxy to galaxy.
\par
This additional factor may be the HI column density; the (qualitative)
agreement between the observed radial profiles of the diffuse fraction
and the predictions of the Dove and Shull models lend support for this
idea.  Dust may also play an important role in regulating the fraction
of Lyc photons which can leak over large distances.  NGC~7793 is known
to have a moderately steep metallicity gradient (Kennicutt, Zaritsky
and Huchra 1994), implying more dust per unit gas mass (and per unit
star formation rate) in the central regions than in the outer parts.
This could result in enhanced absorption of Lyc photons in the central
regions, when compared to the outer disk, and provide at least a
partial explanation for the trend observed in the diffuse fraction.
Furthermore, a higher dust fraction in the central regions would also mean
that less of the DIG produced here would actually be observed.  A
detailed study of the metallicity and extinction in HII regions across
the disks of NGC~7793 and NGC~247 is needed to address this issue
further.

\subsection{Lyc Power Requirements}

The minimum Lyc power requirements we have derived for the DIG in
NGC~247 and NGC~7793 are huge and pose the greatest challenge to all
models for the origin of the DIG.   The recent evolutionary synthesis
models of Leitherer and Heckman (1995) can be used to place constraints
on several possible sources  of the ionizing photons.  These models may
be used to estimate the global star formation rates based on the
observed H$\alpha$ luminosities.  We make the assumption that the DIG
is powered by star formation, and include its contribution in the total
H$\alpha$ luminosities of the galaxies.   Independent estimates of the
SFRs, based on the far--infrared luminosities, may then be used as a
check on how realistic an assumption this is.
\par
The [NII] and extinction--corrected total H$\alpha$ (DIG plus HII regions) luminosities were
measured to  be 4.1$\times$10$^{40}$~erg~s$^{-1}$ for NGC~7793 and
1.5$\times$10$^{40}$~erg~s$^{-1}$ for NGC~247.  Formula (9) of
Leitherer and Heckman (1995) then translates these values into Lyc
photon production rates of $3.03 \times 10^{52}$~s$^{-1}$ and $1.10
\times 10^{52}$~s$^{-1}$.  These photon production rates correspond to
global star formation rates (M$>$ 1M$_{\sun}$) of 0.096 M$_{\sun}$
yr$^{-1}$ for NGC~7793 and 0.035 M$_{\sun}$ yr$^{-1}$ for NGC~247,
assuming a Salpeter IMF with M$_{upper}$ = 100 M$_{\sun}$ and solar
metallicity (Leitherer and Heckman 1995).    It should be noted that
metallicity only influences the derived quantities to a small extent.
Adopting  M$_{lower}$=0.1M$_{\sun}$ would increase these star formation
rates by a factor of 2.5.  These values are all tabulated in Table 2.
\par
The integrated infrared luminosities of the galaxies provide an
independent estimate of the global star formation rates.   Using the
measured IRAS fluxes (Rice {\it et al.} 1988) and assuming the distances
in Table 1, the far infrared luminosities were calculated to be 1.9
$\times$ 10$^{42}$ erg~s$^{-1}$ for NGC~7793 and 4.5 $\times$
10$^{41}$ erg~s$^{-1}$ for NGC~247.
From the derivations of Hunter {\it et al.} (1989) (their equation 14),
these luminosities can be converted into star formation rates of
0.24~E$^{-1}$ M$_{\sun}$ yr$^{-1}$ for NGC~7793 and 0.06~E$^{-1}$
M$_{\sun}$ yr$^{-1}$ for NGC~247, where E is a constant of order unity
measuring  the coupling efficiency between the total power radiated by
dust grains and that included in the IRAS--based measurement of
L$_{IR}$.    While these calculations are based on different stellar
lifetimes than those used in the Leitherer and Heckman models and have
several uncertainties, such as the value of the lowest main--sequence stellar mass
that significantly contributes to L$_{IR}$, they agree with the star
formation rates calculated on the basis of the total H$\alpha$
luminosities to within a factor of 2.  Had we been incorrect to
include the H$\alpha$ luminosity of the DIG in our estimate of the star
formation rate, we could have expected much larger discrepancies
between the rates calculated based on these different methods.  Thus,
it seems quite reasonable that the DIG is indeed powered by Lyc photons
from OB stars.
\par
We can use the set of Leitherer and Heckman (1995) models which are
based on a constant SFR to estimate the mechanical energy that is
injected into the ISM  from Type~II supernovae and stellar winds.
These models give a mechanical luminosity of $7.6 \times
10^{40}$~erg~s$^{-1}$ for NGC~7793 and $2.7 \times
10^{40}$~erg~s$^{-1}$ for NGC~247, based on the assumptions mentioned
above and using the star formation rates calculated from the H$\alpha$
luminosities (note that these mechanical outputs would be increased by
a factor of two if we assumed a Miller--Scalo IMF).  These numbers can
be compared with the minimum Lyc power required to ionize the DIG,
derived in Section 4.4; they fall short by a factor of 2.5 in NGC~7793
and by a factor of 4.4 in NGC~247.   If we compare them to the more
realistic power requirements, calculated on the basis of the {\it
corrected} DIG H$\alpha$ luminosities,  then they fall short by factors
of 3.6 in NGC~7793 and 4.5 in NGC~247.  Type~I supernovae will provide
an additional  contribution to the mechanical luminosity, but given the
estimates of the relative rates of Type~I and Type~II supernovae in
typical disk galaxies (van den Bergh and Tammann 1992), it is unlikely
that this will change the available energy by more than a factor of
two.  We can thus conclude that  supernovae and stellar winds are
unable to produce a substantial fraction of the DIG ionization in these
galaxies, even assuming that they can transfer their energy to the ISM
with 100\% efficiency.  This result contrasts with that found for the
solar neighborhood  where the energy from supernovae can almost account
for the observed level of DIG ionization (Reynolds 1984).
\par
 Despite the fact that OB stars are the only known source of ionizing
photons which can easily meet the overall power requirements for the
DIG,  several unresolved issues remain.  As pointed out by Reynolds
(1995), observations of HI clouds suggest that there are 3--4 clouds
with N$_{HI}$ $\sim$~3$\times$ 10$^{19}$ cm$^{-2}$ along every 300~pc
line of sight near the midplane of the Galactic disk, with the number
possibly increasing dramatically towards lower column densities
(Kulkarni and Heiles 1987; Dickey and Garwood 1989).  Such clouds are
expected to present a significant opacity to Lyc photons and should
severely limit the distance which photons can travel radially within
disks similar to that of our Galaxy.  In addition, the recent
non-detection of the HeI $\lambda$5876 recombination line in the
direction of relatively high Galactic DIG surface brightness implies
that the local DIG ionization is due to a significantly softer spectrum
than that expected from the bulk of the O~star population in the solar
neighborhood (Reynolds and Tufte 1995).  This result implies that
either stars of spectral type O8 and later provide most of the ionizing
photons for the local DIG, or that some mechanism is in place which
effectively softens the radiation from early type O stars.
\par
Late-type O and early B stars account for 24\% of the total Lyc photons
produced in the solar neighborhood (Vacca, Garmany and Shull 1995),
although this number may be an underestimate (e.g. Cassinelli {\it et
al.} 1995). Assuming this value, and using the Leitherer and Heckman
models to predict the total output in Lyc photons for the inferred star
formation rates, we can estimate the amount of ionizing luminosity
produced by such stars in NGC~7793 and NGC~247.  The Leitherer and
Heckman models predict total ionizing outputs of
$9.6~\times~10^{41}$~erg~s$^{-1}$ in NGC~7793 and
$3.4~\times~10^{41}$~erg~s$^{-1}$ in NGC~247.   The  power from soft
Lyc photons produced directly by late-type O and early B stars is then
estimated to be $2.3~\times~10^{41}$~erg~s$^{-1}$ in NGC~7793 and
$8.2~\times~10^{40}$~erg~s$^{-1}$ in NGC~247.  Comparing these numbers
with Lyc power requirements in each galaxy, we find that the minimum
power requirements can be met in NGC~7793 but not in NGC~247.  Both
numbers fall short of satisfying the more realistic, corrected Lyc
power requirements.   Furthermore, these estimates of the soft Lyc
power available should be regarded as upper limits since a considerable
fraction of the power is consumed in the immediate vicinity of OB stars
and hence is not available to ionize ISM over large scales.  Our simple
calculations indicate that such stars are not likely to produce enough
direct soft Lyc radiation to match the power requirements and imply
that another mechanism may indeed be needed to produce the soft
radiation field which a low HeI$\lambda$5876/H$\alpha$ ratio requires.
The absorption and re--emission of Lyc photons as they travel through
the ISM could have such an effect; possible sites for this process
include `chimney' walls (Norman 1991) and the extended envelopes of HII
regions.  Detailed maps of the ionization structure in the HII region --
DIG transition zone would be required to  investigate
these issues more thoroughly and to place tighter constraints on the
actual soft Lyc power that is available.  
\par
Of course, as mentioned above, there are many known sources of Lyc
photons, and they all must contribute at some level to produce the
widespread ionization of the DIG which is observed.

\section{Implications For Measuring Star Formation Rates}

Massive star formation in galaxies is commonly traced by H$\alpha$
emission from HII regions.  A crucial issue is whether the contribution
of the DIG should be included in this calculation.  We have concluded
that the large scale radial distribution and intensity of the DIG
across NGC 7793 and NGC~247, coupled with the global power
requirements, strongly support a picture where the DIG is ionized
predominantly by Lyc photons which have leaked from sites of recent
star formation.  In this scenario, the H$\alpha$ emission from the DIG is as
much related to present--day massive star formation as is the H$\alpha$
emission detected in discrete HII regions.  Failure to count DIG
photons when estimating the total H$\alpha$ emission then leads to an
underestimate both of the global and radial variation of SFRs in galaxies, especially in
the outermost parts where the DIG contribution is more substantial.
For example, had only HII region fluxes been used to calculate the
global star formation rates presented in Section 5.3, then the  derived
rates would have been  29\% lower for NGC~7793 and 50\% lower for
NGC~247.
\par
Most integrated measurements of global SFRs are based on large-aperture
H$\alpha$ photometry, and hence both the HII region and DIG
contributions are automatically included.  On the other hand,  it has
been common practice to derive the radial variation of the SFR based on
counts of discrete HII regions alone (eg.  Kennicutt 1989).  In this
case, the exclusion of the DIG component becomes a more serious issue
and particularly affects the star formation rates derived for the outer
parts of galaxies, where the DIG contribution to the H$\alpha$ emission
is considerable.  Kennicutt {\it et al.} (1995) also noted the possible
importance of counting the DIG in deriving radial star formation rates
and raised the crucial issue of how to estimate the location of the
origin of the DIG photons.   If significant Lyc photon transport occurs
within disks, then the location of the recombination photon (traced by
H$\alpha$ emission) may be very different from the location where the
Lyc photon originated.   Our data are binned in radial bins of size
350~pc (NGC~7793) and 500~pc (NGC~247), typical of the distance which
is predicted to trap  $>$ 90\% of the escaping Lyc  photons vertically
in the disk of our Galaxy (Dove and Shull 1994). Since 350~pc
corresponds to  $\sim$~two HI scaleheights in our Galaxy and is a small
fraction of an HI scalelength, this suggests that possibly less
transport occurs over a similar distance radially within the disk.
 Thus we expect that less than 10\% of the DIG photons in a given bin
will have been radially transported from other locations in the disk.
As noted earlier, DIG can generally be found at mean distances of
$\sim$~500~pc from bright  HII regions; this provides another
constraint on the typical scales over which radial transport takes
place and is in good agreement with the distance required to produce
the appropriate dilution of the Lyc flux from O stars to explain the
observed DIG line ratios (eg. Reynolds 1994).   Constraints on the
amount of radial photon transport can come from calculating the total
Lyc power required per annulus as a function of radius.  Both galaxies
have almost constant power requirements per annulus across their disks,
declining significantly only in the last few bins.  Thus, if radial
transport were occurring over significant scales, a very large,
probably unrealistic fraction of the total Lyc photons would have to be
transported.

\begin{figure}[h]
\figurenum{7}
\plottwo{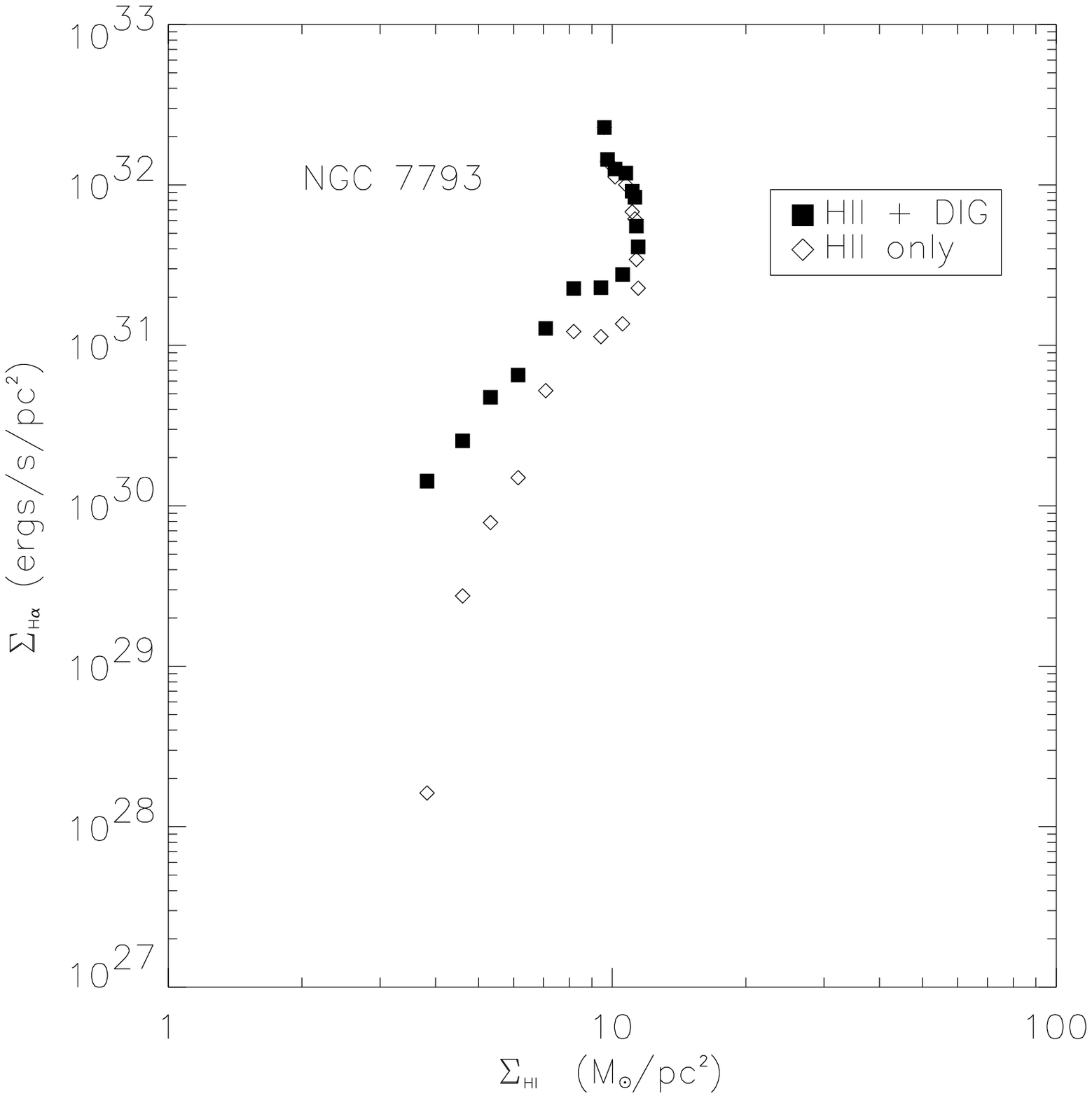}{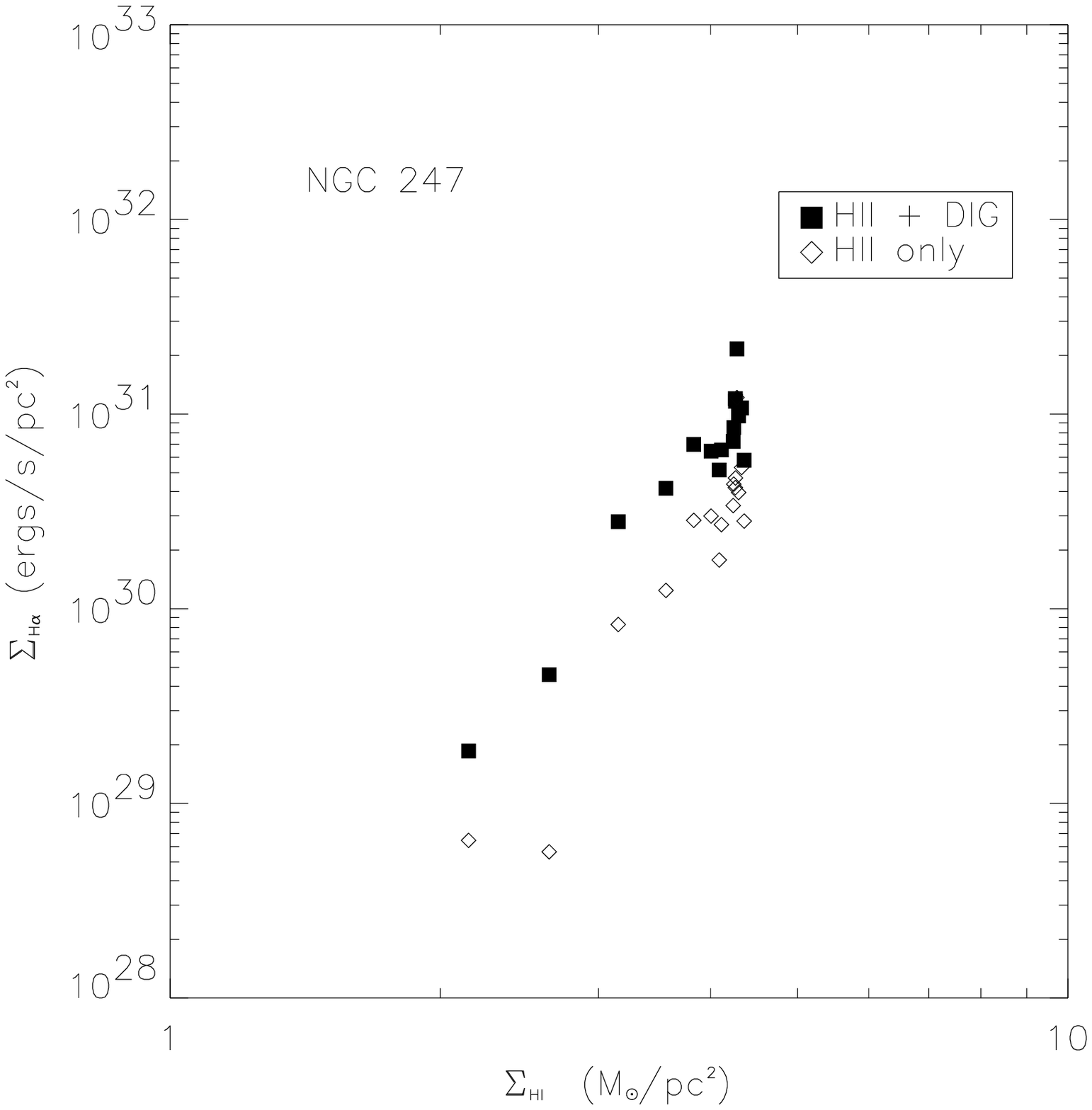}
\caption{(a) Plot of the H$\alpha$ surface brightness
(proportional to SFR/area), derived on the basis of counting both total
H$\alpha$ emission (filled squares) and HII region emission alone (open diamonds),
versus the HI surface density for NGC~7793.  (b)  Same as in (a) for
NGC~247.}
\end{figure}

Figures 7a and b illustrate the variation of H$\alpha$ surface brightness
(proportional to SFR/area), derived on the basis of counting either
total H$\alpha$ emission or only HII region emission, versus the  HI
surface density for each galaxy (Carignan \& Puche 1990a; 1990b).  Such
plots provide information on the physics underlying the star formation
law in galaxies.  The contribution of molecular hydrogen is not
included in these plots.  The two estimates of the SFR/area show very
different behaviors.  The estimates based on HII regions alone show a
sharp cut-off at low HI column densities.  Such an effect has been
interpreted as being the result of a threshold for massive star
formation (Kennicutt 1989).  This evidence for an abrupt cut-off is
{\it not} seen in our data when the DIG component is included in
determining the star formation rate.
\par
 A similar result was found by Ryder and Dopita (1993) in their study of
radial surface brightness profiles in a large sample of southern
spirals.  They noted that H$\alpha$ scalelengths are on average larger
than V and I scalelengths when areal surface photometry is carried out.
This result contrasts with that found by Kennicutt (1989), using HII
region counts alone, who found that broadband (BVR) scalelengths are
generally very similar to H$\alpha$ scalelengths in galactic disks.
Ryder and Dopita (1993) have suggested this discrepancy may be due to
the fact that areal surface photometry takes into account a faint,
diffuse component of H$\alpha$ emission which could play an important
role in the outer disks of galaxies.  In this paper, we find compelling
evidence that this is indeed the case.

\section{Summary}

We have presented a study of the large scale distribution and global
energetics of widespread diffuse ionized gas lying outside the
boundaries of discrete HII regions, identified from deep H$\alpha$
images of the nearby Sculptor spirals NGC~247 and NGC~7793.   We have
separated the total H$\alpha$ emission into that from discrete HII
regions and that from diffuse emission by means of an isophotal cut in
surface brightness, corresponding to an emission measure of 80 pc
cm$^{-6}$.  Most of the H$\alpha$ emission lying below this value is
clearly diffuse and/or filamentary.
\par
Radial and azimuthal intensity distributions of the DIG reveal that it
is highly correlated with bright HII regions over both small and large
scales.  On the scales of a few hundred parsecs, the bright DIG is
localised around individual luminous HII regions in frothy, filamentary
halos.  At faint surface brightnesses, the DIG is ubiquitous, though
maintains the overall pattern and extent of the bright star--forming
disk.    Over larger scales, the DIG has spatial and intensity
distributions which closely follow those of the discrete HII regions,
and, hence, of the mean level of star formation.
\par
The observed DIG H$\alpha$ luminosities are considerable and the DIG
contributes 50\% of the measured total H$\alpha$ luminosity in NGC~247
and 29\% in NGC~7793.   These values are lower limits derived by assuming
that the intensity of the DIG overlying discrete HII regions is
zero.  More realistic
values of the diffuse fractions, calculated assuming that DIG fills the
entire area of the elliptical annuli used, are 53\% in NGC~247 and 41\%
in NGC~7793.  These values are remarkably similar to the global diffuse
fractions found in other actively star--forming galaxies, which span a
range of morphological types and star formation rates.   This
unexpected result suggests some global regulation of the fraction of
photons which can leak from HII regions and are available to ionize the
ISM over large scales.  The radial variation of the diffuse fraction
shows an increase across the disk of NGC~7793 and is relatively
constant across the disk of NGC~247, despite the declining star
formation rate per unit area (ie. Lyc production rate) with increasing
radius.  This behaviour suggests that the additional factor which
regulates photon leakage from HII regions is one which has some radial
dependence.  The qualitative agreement between our observations and the
predictions of the Dove and Shull (1995) models for photoionization of
the DIG suggest that this factor may be the HI column density.  We note
the role that dust might play in regulating photon leakage in
galaxies.  More detailed metallicity gradients are required for these
galaxies in order to further investigate this issue.
\par
The integrated miminum Lyc powers required to sustain the DIG in these
galaxies are enormous and place tight constraints on the source of the
ionizing photons.  Only massive star formation can easily satisfy the
power requirements, with the mechanical luminosity from supernovae and
stellar winds falling short by more than a factor of two.  This result
contrasts with that found for the local Galactic DIG where the energy
from supernovae can almost account for the observed level of
ionization.  In view of the new constraints imposed on the origin of
the DIG photons from the non--detection of the HeI$\lambda$5876
recombination line in the Galaxy, we estimated the Lyc power supplied
by stars of spectral type O8 and later.   The available power falls
short of realistic estimates of the required Lyc power, implying that 
direct radiation from such stars alone cannot acount for the entire
ionization of the DIG.  This suggests that either some
mechanism is in place which softens the radiation field of early O
stars, which produce the bulk of the Lyc photons, or that the photons
which ionize the DIG come from a variety of sources,  all contributing
at some level to produce the widespread ionization which is observed.
\par
Our results strongly support an interpretation of the DIG as being due
to Lyc photon leakage from discrete HII regions.   In such a scenario,
the  DIG is a direct product of recent massive star formation and its
contribution to the H$\alpha$ emission line flux of the galaxy should
be included in order to derive accurate star formation rates.  We have
illustrated the importance of taking the DIG into account when studying
the correlation between star formation rate per unit area  and HI gas
surface density.  If HII regions alone are used to trace the star
formation rate per unit area, then these plots show a sharp cut--off in
star formation activity  at low HI column densities, commonly
interpreted as being the result of a threshold for massive star
formation (Kennicutt 1989).   This evidence for an abrupt cut--off is
not seen in our data when the DIG component is included in determining
the star formation rate per unit area.

\acknowledgments

We thank Piero Rosati for many useful discussions during the course of
this work, both scientific and technical, and Tim Heckman, Colin Norman
and Gerhardt Meurer for commenting on an earlier draft of this paper.
The anonymous referee provided comments which helped us improve the
presentation of our results.  We thank the staff of Cerro Tololo
Inter--American Observatory for their excellent support.  AMNF
acknowledges receipt of an Amelia Earhart Fellowship awarded by the
Zonta International Foundation.  This research has been supported
in part by NASA grant NAGW--2892.  The Center for Particle Astrophysics
is funded by the NSF.
\pagebreak

\pagebreak

\centerline{\bf FIGURE CAPTIONS}
\bigskip
\noindent{Fig.1 -- (a)  H$\alpha$ continuum--subtracted image of
NGC~7793 displayed at high contrast to show only the cores of bright
HII regions.  North is up and east is to the left.   \\ (b) H$\alpha$
continuum--subtracted image of NGC~7793 displayed at low contrast to
show the diffuse ionized gas emission. \\ (c) H$\alpha$
continuum--subtracted image of NGC~247 displayed to show the cores of
bright HII regions.  North is up and east is to the left. \\ (d)
H$\alpha$ continuum-- subtracted image of NGC 247 displayed to show the
diffuse ionized gas emission.}

\noindent{Fig.2 --(a) Subsection of NGC~7793 illustrating the DIG.
Pixels with EM $>$ 80~pc cm$^{-6}$  have
 been masked out (white).
\\ (b)  As in (a) but for NGC~247}

\noindent{Fig.3 -- (a) Plot of the azimuthal variation of uncorrected
H$\alpha$ +[NII] surface brightness at four radii in NGC~7793 .  The
radii correspond to 0.1R$_{25}$, 0.3R$_{25}$, 0.6R$_{25}$ and
0.9R$_{25}$.  The width of each radial bin is 250~pc and the data are
azimuthally binned into 2{\deg} sectors. \\  (b)  As in (a) but for NGC
247.}

\noindent{Fig.4 -- (a) Deprojected radial profile of the total
H$\alpha$ surface brightness (solid line), the H$\alpha$ surface
brightness derived by counting HII regions alone (dashed--dotted line)
and the H$\alpha$ surface brightness of the DIG (dashed line) in NGC
7793.  The optical radius (R$_{25}$) is indicated. \\ (b)  As in (a) but for
NGC~247.}

\noindent{Fig.5 -- Radial variation of the diffuse fraction for NGC7793
(left) and NGC247 (right).}

\noindent{Fig.6 --  Minimum power per unit area required to maintain
the DIG across the disk of each galaxy.  The dotted line indicates the
value derived for the solar neighborhood (Reynolds 1984).}

\noindent{Fig.7 -- (a) Plot of the H$\alpha$ surface brightness
(proportional to SFR/area), derived on the basis of counting both total
H$\alpha$ emission (filled squares) and HII region emission alone (open diamonds),
versus the HI surface density for NGC~7793.  \\
(b)  Same as in (a) for
NGC~247.}

\end{document}